\documentclass[12pt]{article}
\usepackage{amssymb,amsmath,epsfig}

\begin{document}
\allowdisplaybreaks
\title{\bf Dynamics of Particles Near Black Hole with Higher Dimensions}
\author{M. Sharif \thanks{msharif.math@pu.edu.pk} and
Sehrish Iftikhar
\thanks{sehrish3iftikhar@gmail.com}~~\thanks{On leave from Department of Mathematics, Lahore College
for Women University, Lahore-54000, Pakistan.}\\
Department of Mathematics, University of the Punjab,\\
Quaid-e-Azam Campus, Lahore-54590, Pakistan.}
\date{}
\maketitle
\begin{abstract}
This paper explores the dynamics of particles in higher dimensions.
For this purpose, we discuss some interesting features related to
the motion of particles near Myers-Perry black hole with arbitrary
extra dimensions as well as single non-zero spin parameter. Assuming
it as a supermassive black hole at the center of galaxy, we
calculate red-blue shifts in the equatorial plane for the far away
observer as well as corresponding black hole parameters of the
photons. Next, we study the Penrose process and find that the energy
gain of particle depends on the variation of black hole dimensions.
Finally, we discuss the center of mass energy for eleven dimensions
which indicates similar behavior as that of four dimensions but it
is higher in four dimensions than five or more dimensions. We
conclude that higher dimensions have a great impact on the particle
dynamics.
\end{abstract}

\section{Introduction}

Gravity in more than four dimensions has been the subject of
interest in recent years for a variety of reasons. This leads to
significant features of black holes (BHs) like uniqueness, dynamical
stability, spherical topology and the laws of BH mechanics. It has
been found that laws of BH mechanics are universal while the
properties of BH are dimension dependent. The concept of higher
dimensions became prominent in the 19th century with the
Kaluza-Klein theory which unified gravitation and electromagnetism
in five dimensions \cite{1}. Later on, development of string and
M-theories led to further progress in higher dimensional gravity.
String theory is the most promising candidate of quantum gravity -
the fascinating theory of high energy physics. M-theory is the
generalization of superstring theory that gave the concept of eleven
dimensions. Charged BHs in string theory play an important role in
understanding the BH entropy near extremal limits \cite{2}. Callan
and Maldacena \cite{3} calculated Hawking temperature, radiation
rate and entropy for the extremal Reissner-Nordstr$\ddot{o}$m BH in
the context of string theory and proposed that quantum evolution of
BH does not lead to information loss. Itzhaki et al. \cite{4}
studied D-brane solutions in string theory for the region where
curvature is very small.

The study of BHs in higher dimensions has attracted many
researchers. Tangherlini \cite{5} was the first who generalized the
Schwarzschild BH to arbitrary extra dimensions ($D>4$) while the
Myers and Perry generalized the Kerr BH \cite{6}. There also exist
black rings \cite{7} and multi BH solutions like black Saturns and
multi black rings \cite{8}. Carter and Neupane \cite{9} studied
stability and thermodynamics of higher dimensional Kerr-anti de
Sitter BH and found stability for equal rotation parameters. Dias et
al. \cite{10} investigated perturbation of Myers-Perry (MP) BH and
found stability in five and seven dimensions. Murata \cite{11} found
instabilities of $D$-dimensional MP BH and concluded that there is
no evidence of instability in five dimensions, however, for
$D=7,9,11,13$, the spacetime became unstable due to large angular
momenta.

Galactic rotation curves are based on the measurement of red-blue
shifts of emitted light from distant stars. Due to galaxy rotation,
one side of the galaxy will appear to be blue shifted as it rotates
towards the observer and the other will be red-shifted as it rotates
away from the observer \cite{16}. Nucamendi et al. \cite{17} studied
the rotation curves of galaxies by measuring the frequency shifts of
spherically symmetric spacetime. Lake \cite{18} showed that galactic
potential can be linked to red-blue shifts of the galactic rotation
curves. Bharadwaj and Kar \cite{19} proposed that the flat rotation
curves of the spiral galaxies can be explained by dark matter halos
having anisotropic pressure. Moreover, the deflection of light ray
is sensitive to the pressure of the dark matter. Faber and Visser
\cite{20} argued that observations of galactic rotation curves
together with gravitational lensing describe the deduction of
galactic mass and provides information about the pressure of
galactic fluid. Herrera-Aguilar et al. \cite{21} presented a useful
technique to study red-blue shifts for a spiral galaxy by
generalizing the galactic rotation curves for spherically symmetric
spacetime to an axisymmetric metric. This approach has been used to
express the Kerr BH parameters in terms of red-blue shift functions
\cite{22}.

Penrose process is related to the energy extraction from a rotating
BH which depends upon the conservation of angular momentum.
Chandrasekhar \cite{23} studied the Penrose process for the Kerr BH
and discussed the nature of this process  as well as examined the
limits on the extracted energy. He found that in the equatorial
plane, only retrograde particles possess negative energy and the
particles should remain inside the static limit (ergosphere). Bhat
et al. \cite{24} investigated the Penrose process for the
Kerr-Newman BH and concluded that energy becomes highly negative in
the presence of electromagnetic field while for neutral particles,
the gain energy decreases in the presence of charge of the BH.
Recently, Lasota et al. \cite{25} presented the generalized Penrose
process and stated that ``for any matter or field, tapping the BH
rotation energy is possible if and only if negative energy and
angular momentum are absorbed by BH and no torque at the BH horizon
is necessary (or possible)". There are some other important results
\cite{26} in the context of Penrose process.

The collision energy of particles in the frame of center of mass
results in the formation of new particles and the energy produced in
this process is known
as center of mass energy. The center of mass energy of two colliding
particles is infinitely high near the event horizon of maximally
spinning Kerr BH \cite{27}. This approach is very useful as it
describes rotating BH as a particle accelerator at the Planck energy
scale. Lake \cite{28} examined particle collision for non-extremal
Kerr BH at the inner horizon and found center of mass energy to be
finite. The center of mass energy is also analyzed for the
Kerr-Newman BH which shows the dependence on the spin and charge of
the BH \cite{29}. The same mechanism was employed on the Kerr-Newman
Tuab \cite{30} and rotating Hayward BH \cite{31}. Other important
aspects related to center of mass energy have been explored in
\cite{32}.

In this paper, we study the dynamics of particles for a
$D$-dimensional MP BH in the equatorial plane. The paper is
organized as follows. In the next section, we review timelike
geodesics in higher dimensions. In section \textbf{3}, we study
red-blue shifts of MP BH and formulate BH parameters in terms of
red-blue shift functions. Section \textbf{4} explores the Penrose
process and section \textbf{5} is devoted to study the center of
mass energy for this BH. We conclude our results in the last
section.

\section{Review of Geodesics in Higher Dimensions}

The generalization of Kerr BH in higher dimensions, i.e., $D>4$, is
known as MP BH \cite{6} and shares many properties with Kerr BH.
This plays significant role to explore gravity in higher dimensions
as it provides a new vision about important features of event
horizons. There are several choices of rotation axis as well as
angular momentum regarding particular rotation plane. We consider a
simple case by considering a single spinning parameter $a$. The
$D$-dimensional MP BH in Boyer-Lindquist coordinates is given as
\cite{33,34}
\begin{eqnarray}\nonumber
ds^{2}&=&-(\frac{\Delta-a^2\sin^2\theta}{\rho^2})dt^2
+\frac{\rho^2}{\Delta}dr^2+\rho^2d\theta^2
\\\nonumber
&+&\frac{(r^2+a^2)^2-\Delta a^2\sin^2\theta}{\rho^2}\sin^2\theta
d\phi^2
\\\label{1}
&-&2\frac{(r^2+a^2)-\Delta}{\rho^2}a\sin^2\theta
dtd\phi+r^2\cos^2\theta d\Omega^2_{D-4},
\end{eqnarray}
where
\begin{eqnarray*}
\rho^{2}&=&r^{2}+a^2\cos\theta^{2},\quad \Delta=r^2+a^2-\mu r^{5-D},\\
\mu&=&\frac{16\pi GM}{(D-2)\Omega_{(D-2)}},\quad
\Omega_{D-2}=\frac{2\pi^{\frac{D-1}{2}}}{\Gamma(\frac{D-1}{2})},
\end{eqnarray*}
and
\begin{equation*}
d\Omega^{2}_{D-2}=d\theta^{2}_{1}+\sin^{2}\theta_{1}d\theta^{2}_{2}+
\sin^{2}\theta_{1}\sin^{2}\theta_{2}d\theta^{2}_{3}+...+
\prod^{D-3}_{\lambda=1}\sin^{2}\theta_{\lambda}d\theta^{2}_{D-2},
\end{equation*}
describes the $(D-4)$ unit sphere. This metric is asymptotically
flat, vacuum spacetime with ADM mass $\mu$ and $D$ may be even or
odd. For $D=4$ it reduces to the Kerr BH while $a=0$ leads to the
Schwarzschild BH. The event horizon of (\ref{1}) is the largest root
of $\Delta=0$
\begin{equation}\nonumber
r^2_{h}+a^2-\mu r^{5-D}_{h}=0.
\end{equation}
The extremal limit exists for $D=4,5$ ($a<\frac{\mu}{2}$ and
$a<\sqrt{\mu}$). For $D\geq6$, there is only one positive root when
$a>0$, which indicates that there is no extremal limit in higher
dimensions \cite{34}. The behavior of horizons along the spin
parameter can be seen in Figure \textbf{1}.
\begin{figure}\centering
\epsfig{file=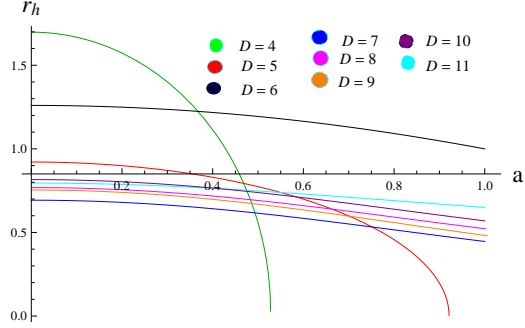,width=.50\linewidth} \caption{Horizons with
respect to $a$.}
\end{figure}
The particle motion can be described by the Lagrangian
\begin{equation}\label{r}
\mathcal{L}=\frac{1}{2}g_{\nu\sigma}\dot{x}^{\nu}\dot{x}^{\sigma},
\end{equation}
where $\dot{x}^{\nu}=u^{\nu}=dx^{\nu}/d\tau$ and $u^{\nu}$ is the
particle's $D$-velocity and $\tau$ is the affine parameter. In
equatorial plane ($\theta=\frac{\pi}{2},~\dot{\theta}=0$),
Eq.(\ref{r}) takes the form
\begin{eqnarray}\label{3}
2\mathcal{L}=-(1-\frac{\mu}{r^{D-3}})\dot{t}^2
-2a(\frac{\mu}{r^{D-3}})\dot{t}\dot{\phi}
+\frac{r^2}{\Delta}\dot{r}^2+ (r^2+a^2+\frac{\mu
a^2}{r^{D-3}})\dot{\phi}^2.
\end{eqnarray}
The generalized momenta for (\ref{1}) are calculated as
\begin{eqnarray}\label{9}
-k_{t}&=&-(1-\frac{\mu}{r^{D-3}})\dot{t}
-(\frac{a\mu}{r^{D-3}})\dot{\phi}=E,
\\\label{10}
k_{\phi}&=&-(\frac{a\mu}{r^{D-3}})\dot{t}+(r^2+a^2+\frac{\mu
a^2}{r^{D-3}})\dot{\phi}=L,
\\\label{11}
k_{r}&=&\frac{r^2}{\Delta}\dot{r},
\end{eqnarray}
where dot represents derivative with respect to $\tau$. We find that
the Lagrangian is independent of $t$ and $\phi$, therefore $k_{t}$
and $k_{\phi}$ are conserved and hence describes stationary and
axisymmetric characteristics of MP BH.

The Hamiltonian can be written as
\begin{eqnarray}\label{12}
H=k_{t}\dot{t}+k_{\phi}\dot{\phi}+k_{r}\dot{r}-\mathcal{L}.
\end{eqnarray}
For the metric (\ref{1}), it takes the following form
\begin{eqnarray}\nonumber
2H&=&-[(1-\frac{\mu}{r^{D-3}})\dot{t}
+(\frac{a\mu}{r^{D-3}})\dot{\phi}]\dot{t}
+[-(\frac{a\mu}{r^{D-3}})\dot{t}+(r^2+a^2+\frac{\mu
a^2}{r^{D-3}})\dot{\phi}]\dot{\phi}+\frac{r^2}{\Delta}\dot{r}^2
\\\label{13}
&=&-E\dot{t}+L\dot{\phi}+\frac{r^2}{\Delta}\dot{r}^2=\delta=\text{constant},
\end{eqnarray}
where $\delta=0,-1,1$ describe null (lightlike), timelike and
spacelike geodesics. From Eqs.(\ref{9}) and (\ref{10}), we obtain
\begin{eqnarray}\label{14}
\dot{t}&=&\frac{1}{\Delta}
[\frac{a\mu}{r^{D-3}}E+(1-\frac{a\mu}{r^{D-3}})L],
\\\label{15}
\dot{\phi}&=&\frac{1}{\Delta}[(r^2+a^2+\frac{\mu
a^2}{r^{D-3}})E-\frac{a\mu}{r^{D-3}}L].
\end{eqnarray}
Inserting Eqs.(\ref{14}) and (\ref{15}) into (\ref{13}), we find
radial equation of motion
\begin{equation}\label{17}
r^2\dot{r}^2=r^2E^2+\frac{\mu}{r^{D-3}}(aE-L)^2+(a^2E^2-L^2)+\Delta
\delta.
\end{equation}
Equations (\ref{14})-(\ref{17}) are very important as they can be
used to study various features related to particle motion around
(\ref{1}). Following some algebraic manipulation, the energy and
angular momentum can be written as
\begin{eqnarray}\nonumber
E&=&\frac{1}{\sqrt{\zeta_{\pm}}}[1-\mu y^{D-3}\mp
a\sqrt{(D-3)(\mu/2) y^{D-1}}],
\\\label{18}
L&=&\mp \frac{\sqrt{(D-3)(\mu/2)y^{D-4}}}{y\zeta_{\mp}}[1+a^2y^2\pm
\sqrt{\frac{(\mu/2) y^{D-1}}{D-3}}],
\end{eqnarray}
where $y=1/r$ and $\zeta_{\pm}=1-(D-1)(\mu/2)y^{D-3}\pm
2a\sqrt{(D-3)(\mu/2)y^{D-1}}$. These are the same as obtained in
\cite{33} for $M=\mu/2$ ($M$ is a parameter related to the BH mass).
Here, we do not consider this substitution as we are interested in
finding our results with ADM mass.

\section{Red-Blue Shifts of Myers-Perry BH}

This section is devoted to study red-blue shifts in higher
dimensions for MP BH. Herrera-Aguilar et al \cite{21} discussed
red-blue shifts for an axially symmetric spacetime and presented a
convenient approach to study galactic rotational curves of the
spiral galaxies. Since spiral galaxies possess axial symmetry, so
this method provides an information about the interior of the
gravitational field of such galaxies. Herrera-Aguilar \cite{22}
extended this technique for the Kerr BH and found parameters in
terms of red-blue shifts in the equatorial plane. Following
\cite{21,22}, we generalize these results for the MP BH. We consider
two observers $O_{d}$ and $O_{e}$ which correspond to detector and
light emitter (star) placed at points $P_{d}$ and $P_{e}$,
respectively. The detector and emitter possess $D$-velocities
$u^{\nu}_{d}$ and $u^{\nu}_{e}$. We assume that the stars are moving
in the galactic plane such that polar angle is fixed
($\theta=\frac{\pi}{2}$). In this case, we have
$u^{\nu}_{e}=(u^{t},~u^{r},0,...,0,~u^{\phi})_{e}$, where
$u^{\nu}=dx^{\nu}/d\tau$ and $\tau$ is the proper time of the
particle. The $D$-velocity of the detector,
$u_{d}=(u^{t},~u^{r},0,...,0,~u^{\phi})_{d}$, is located far away
from the source. The component $u^{\phi}$ is related to the
observer's dragging at point $P_{d}$ due to galactic rotation and
its effect is present when measuring red-blue shifts in our galaxy
(Milky Way) or nearby galaxies \cite{21}.

The general frequency expression of photon measured by an observer
is
\begin{equation}\label{18a}
\omega_{c}=-k_{\nu}u^{\nu}_{c}|_{P_{c}},
\end{equation}
where $k^{\nu}=(k^{t},k^{r},0,...,0,k^{\phi})_{c}$ is the
$D$-momentum in the equatorial plane and index $c$ corresponds to
emitter ($e$) or detector ($d$) for the spacetime at point $P_{c}$.
The light frequencies detected by an observer at $P_{d}$ and
measured by an observer moving with the emitted particle at point
$P_{e}$ are
\begin{equation}\label{19}
\omega_{d}=-k_{\nu}u^{\nu}_{d},\quad \omega_{e}=-k_{\nu}u^{\nu}_{e}.
\end{equation}
The frequency shift corresponding to emission as well detection has
the form
\begin{eqnarray}\label{20}
1+z=\frac{\omega_{e}}{\omega_{d}}
=\frac{(Eu^{t}-Lu^{\phi}-g_{rr}u^{r}k^{r})|_{e}}{(Eu^{t}-Lu^{\phi}
-g_{rr}u^{r}k^{r})|_{d}}.
\end{eqnarray}
Since we have considered circular orbital motion ($u^{r}=0$), the
above equation becomes
\begin{eqnarray}\label{21}
1+z=\frac{(Eu^{t}-Lu^{\phi})|_{e}}{(Eu^{t}-Lu^{\phi})|_{d}}=
\frac{u^{t}_{e}-b_{e}u^{\phi}_{e}}{u^{t}_{d}-b_{d}u^{\phi}_{d}},
\end{eqnarray}
where $b=\frac{L}{E}$ is the impact parameter for the observer
located at infinity. This parameter is zero when it is measured from
either side of the center of galaxy. In this case, Eq.(\ref{21}) can
be written as
\begin{equation}\label{22}
1+z= \frac{u^{t}_{e}}{u^{t}_{d}}.
\end{equation}
Subtracting (\ref{22}) from (\ref{21}), the kinematical red shift
can be obtained as
\begin{equation}\nonumber
z_{\kappa}=z-z_{c}=
\frac{u^{t}_{e}u^{\phi}_{d}b_{d}-u^{t}_{d}u^{\phi}_{e}b_{e}}{u^{t}_{d}
(u^{t}_{d}-b_{d}u^{\phi}_{d})}.
\end{equation}

It is important to mention here that the impact parameter remains
constant along the path of photon, i.e. $b_{e}=b_{d}$. This is due
to the fact that energy and angular momentum are preserved from
emission to detection along the null geodesics. We are interested in
red-blue shifts from either side of the galactic center which
requires two values of $b$ to calculate red-blue shifts. The impact
parameter can be calculated from radial null geodesic
$k^{\nu}k_{\nu}=0$ \cite{21}
\begin{equation}\label{22a}
b_{\pm}=\frac{-g_{t\phi} \pm
\sqrt{g^{2}_{t\phi}-g_{tt}g_{\phi\phi}}}{g_{tt}},
\end{equation}
where $b_{-}$ and $b_{+}$ lead to the red-blue shifts of the photons
emitted from an object moving away or approaching towards a far away
observer located at infinity \cite{21}
\begin{eqnarray}\label{23}
z_{red}=\frac{u^{t}_{e}u^{\phi}_{d}b_{d_{-}}-u^{t}_{d}u^{\phi}_{e}
b_{e_{-}}}{u^{t}_{d}(u^{t}_{d}-u^{\phi}_{d}b_{d_{-}})},
\\\label{23a}
z_{blue}=\frac{u^{t}_{e}u^{\phi}_{d}b_{d_{+}}-u^{t}_{d}u^{\phi}_{e}
b_{e_{+}}}{u^{t}_{d}(u^{t}_{d}-u^{\phi}_{d}b_{d_{+}})},
\end{eqnarray}
where $z_{red}\neq z_{blue}$, in general. The angular velocity can
be defined as
\begin{equation}\label{24}
\frac{u^{\phi}_{d}}{u^{t}_{d}}=\frac{d \phi}{dt} \equiv \Omega_{d}.
\end{equation}
When the observer is located far away from the source (emitter) such
that $u^{\phi}_{d}<<u^{t}_{d}$, then $\Omega_{d}<<1$. Using
Eq.(\ref{24}), Eqs.(\ref{23}) and (\ref{23a}) take the form
\begin{eqnarray}\label{23aa}
z_{red}=\frac{u^{t}_{e}\Omega_{d} b_{d_{-}}-u^{\phi}_{e}b_{e_{-}}}
{u^{t}_{d}(1-\Omega_{d}b_{d_{-}})},
\\\label{23aaa}
z_{blue}=\frac{u^{t}_{e}\Omega_{d} b_{d_{+}}-u^{\phi}_{e}b_{e_{+}}}
{u^{t}_{d}(1-\Omega_{d}b_{d_{+}})}.
\end{eqnarray}

In order to calculate red-blue shifts for the MP BH, we consider
$D$-velocity components corresponding to (\ref{14}) and (\ref{15})
for the circular orbits in the equatorial plane
\begin{eqnarray}\label{255}
u^{t}&=&\frac{1}{\Delta}[\frac{(a\mu)E-(r^{D-3}-\mu)L}{r^{D-3}}],
\\\label{25a}
u^{\phi}&=&\frac{1}{\Delta}[(\frac{r^{D-1}+a^2r^{D-3}+a^2\mu)E-(a\mu)L}{r^{D-3}}],
\end{eqnarray}
where
\begin{eqnarray}\label{266}
E&=&\frac{r^{\frac{D-1}{2}}-\mu r^{\frac{5-D}{2}}\mp a
[(D-3)(\mu/2)]^{\frac{1}{2}}}
{r^{\frac{D-1}{4}}[r^{\frac{D-1}{2}}-(D-1)(\mu/2)r^{\frac{5-D}{2}}\pm
2a[(D-3)(\mu/2)]^{\frac{1}{2}}]^{\frac{1}{2}}},\\\label{26a}
L&=&\frac{\pm ((D-3)(\mu/2))^{\frac{1}{2}}[r^2+a^2 \mp
2a(\frac{(\mu/2)}{D-3})^{\frac{1}{2}}r^{\frac{5-D}{2}}]}
{r^{\frac{D-1}{4}}[r^{\frac{D-1}{2}}-(D-1)(\mu/2)r^{\frac{5-D}{2}}\pm
2a[(D-3)(\mu/2)]^{\frac{1}{2}}]^{\frac{1}{2}}},
\end{eqnarray}
here $\pm$ sign describe direct and retrograde objects which may be
emitter or detector with respect to angular velocity. Using
Eqs.(\ref{266}) and (\ref{26a}) in (\ref{255}) and (\ref{25a}), we
obtain
\begin{eqnarray}\label{277}
u^{\phi}&=&\frac{\pm [(D-3)(\mu/2)]^{\frac{1}{2}}}
{r^{\frac{D-1}{4}}[r^{\frac{D-1}{2}}-(D-1)(\mu/2)r^{\frac{5-D}{2}}\pm
2a[(D-3)(\mu/2)]^{\frac{1}{2}}]^{\frac{1}{2}}},\\\label{27a}
u^{t}&=&\frac{\pm [(D-3)(\mu/2)]^{\frac{1}{2}}[r^{\frac{D-1}{2}}\pm
((D-3)(\mu/2))^{\frac{1}{2}}]}
{r^{\frac{D-1}{4}}[r^{\frac{D-1}{2}}-(D-1)(\mu/2)r^{\frac{5-D}{2}}\pm
2a[(D-3)(\mu/2)]^{\frac{1}{2}}]^{\frac{1}{2}}}.
\end{eqnarray}
The angular velocity of the orbiting source (emitter or detector)
around the MP BH can be easily calculated from Eqs.(\ref{277}) and
(\ref{27a}) as
\begin{equation}
\Omega_{\pm}=\frac{\pm
[(D-3)(\mu/2)]^{\frac{1}{2}}}{r^{\frac{D-1}{2}}\pm
[(D-3)(\mu/2)]^{\frac{1}{2}}}.
\end{equation}
The impact parameter for the equatorial circular orbits in
$D$-dimensions can be obtained from Eq.(\ref{22a})
\begin{equation}
b_{\pm}=\frac{-a\mu \pm r^{D-3}[r^2+a^2-\mu
r^{5-D}]^{\frac{1}{2}}}{r^{D-3}-\mu}.
\end{equation}

The red-blue shifts for (\ref{1}) are found from Eqs.(\ref{23aa})
and (\ref{23aaa}) as
\begin{eqnarray}\nonumber
z_{red}&=& \frac{\pm\sqrt{(D-3)(\mu/2)}}{\beta
\gamma(r^{\frac{D-1}{2}}_{d}\pm a\sqrt{(D-3)(\mu/2)})}
\\\label{23d}
&\times&\alpha(r^{\frac{{D-1}}{2}}_{d}-r^{\frac{{D-1}}{2}}_{e})
[a\mu+r^{D-3}_{e}\sqrt{r^2_{e}+a^2-\mu r^{D-3}_{e}}],
\\\nonumber
z_{blue}&=&\frac{\pm \sqrt{(D-3)(\mu/2)}}{\beta
\varrho(r^{\frac{D-1}{2}}_{d}\pm a\sqrt{(D-3)(\mu/2)})}
\\\label{23dd}
&\times&\alpha(r^{\frac{{D-1}}{2}}_{d}-r^{\frac{{D-1}}{2}}_{e})
[a\mu-r^{D-3}_{e}\sqrt{r^2_{e}+a^2-\mu r^{D-3}_{e}}] ,
\end{eqnarray}
where $r_{d}$ and $r_{e}$ are the orbit radius of detector as well
as emitter and
\begin{eqnarray}\nonumber
\alpha&=&r^{\frac{{D-1}}{4}}_{d}[r^{\frac{{D-1}}{2}}_{d}-
(D-1)(\mu/2)r^{\frac{{5-D}}{2}}_{d}\pm2a\sqrt{(D-3)(\mu/2)}]^{\frac{1}{2}},
\\\nonumber
\beta&=&r^{\frac{{D-1}}{4}}_{e}[r^{\frac{{D-1}}{2}}_{e}-
(D-1)(\mu/2)r^{\frac{{5-D}}{2}}_{e}\pm2a\sqrt{(D-3)(\mu/2)}]^{\frac{1}{2}},
\\\nonumber
\gamma &=&r^{\frac{D-1}{2}}_{d}(r^{D-3}_{e}-\mu)\pm
ar^{D-3}_{e}\sqrt{(D-3)(\mu/2)}
\\\nonumber
&\pm& \sqrt{(D-3)(\mu/2)}r^{D-3}_{e}\sqrt{r^2_{e}+a^2-\mu
r^{D-3}_{e}},
\\\nonumber
\varrho &=&r^{\frac{D-1}{2}}_{d}(r^{D-3}_{e}-\mu)\pm
ar^{D-3}_{e}\sqrt{(D-3)(\mu/2)}
\\\nonumber
&\mp& \sqrt{(D-3)(\mu/2)}r^{D-3}_{e}\sqrt{r^2_{e}+a^2-\mu
r^{D-3}_{e}}.
\end{eqnarray}
The relation $b_{e}=b_{d}$ yields the equation relating the radius
of emitter and detector
\begin{eqnarray}
r^{2(D-2)}_{d}-\mu r^{D-1}_{d}-r^{2(D-3)}_{d}(b^2_{e}-a^2)
+2b_{e}\mu(b_{e}-a)r^{D-3}_{d}-(\mu)^2(b^2_{e}-a^2)^2=0.
\end{eqnarray}
When detector is far away from the source and
$r_{d}>>\mu\geq a$, the red-blue shifts reduce to
\begin{eqnarray}\label{24f}
&&z_{red}=\frac{\pm \sqrt{(D-3)(\mu/2)}
[a\mu+r^{D-3}_{e}\sqrt{r^2_{e}+a^2-\mu r^{D-3}_{e}}]} {\beta
(r^{D-3}_{e}-\mu)},
\\\label{24ff}
&&z_{blue}=\frac{\pm \sqrt{(D-3)(\mu/2)}
[a\mu-r^{D-3}_{e}\sqrt{r^2_{e}+a^2-\mu r^{D-3}_{e}}]} {\beta
(r^{D-3}_{e}-\mu)}.
\end{eqnarray}
From Eqs.(\ref{24f}) and (\ref{24ff}), the parameters of MP BH can
be expressed in terms of red-blue shifted of the photons emitted
from the source. The spin parameter in terms of red-blue shifts is
given as
\begin{equation}\label{rr}
a^2=\frac{r^{3(D-3)}_{e}(r^{5-D}_{e}-\mu)(z_{red}+z_{blue})^2}
{\mu^2(z_{red}-z_{blue})^2-r^{2(D-3)}_{e}(z_{red}+z_{blue})^2}.
\end{equation}
The mass parameter corresponding to red-blue shifts can be obtained
from the following expression
\begin{eqnarray}\nonumber
&&2(D-3)r^{4D-14}_{e}\mu(z_{red}+z_{blue})^2(r^{5-D}_{e}-\mu)
(r^{D-3}_{e}-\mu)^4
\\\nonumber
&&[\mu^2(z_{red}-z_{blue})^2-r^{2(D-3)}_{e}(z_{red}+z_{blue})^2]
\\\nonumber
&&=[2\mu^3r^{3D-11}_{e}(D-3)(r^{5-D}_{e}-\mu)
-(\mu^2(z_{red}-z_{blue})^2
\\\label{rrr}
&&-r^{2(D-3)}_{e}
(z_{red}+z_{blue})^2)\times(r^{D-3}_{e}-(D-1)(\mu/2))(r^{D-3}_{e}-\mu)^2]^2.
\end{eqnarray}
Equations (\ref{rr}) and (\ref{rrr}) reduce to the results of Kerr
BH when $D=4$. These equations may provide a useful model for the
researchers interested in higher dimensions using experimental data.
One can observe easily how the photons emitted from the source (e.g.
stars) can be red or blue shifted near the higher dimensional BH.
The mass as well as the rotation of MP BH also affects the emitted
photons. Another interesting phenomenon caused by the BH rotation is
Penrose process that takes place into the ergosphere. In the next
section, we examine the particle motion and its consequences as it
enters into the ergosphere.

\subsection{The Penrose Process}

The rotation of spinning BH corresponds to a reservoir of usable
energy and is related to the properties of particles (e.g. photons)
inside the ergosphere (the stationary limit surface) where the
particles can orbit with total negative energy with respect to a
distant observer. The orbits with negative energy have negative
angular momentum with respect to the BH. Non-rotating BHs do not
have such stationary limit surface, however rotating BHs (e.g. Kerr
BH) have such surfaces, called ergosphere. The particles orbiting
with negative energy can exchange energy with other particles. It
was first pointed out by Penrose that this process can be used to
extract energy from the spinning BH. Following \cite{23}, we study
the Penrose process in higher dimensions for the MP BH. Equation
(\ref{17}) gives
\begin{eqnarray}\nonumber
&&E^2[r^{D-3}(r^2+a^2)+\mu a^2]-2a\mu EL -L^2(r^{D-3}-\mu)+
r^{D-3}\delta\Delta=0.
\end{eqnarray}
Solving this equation for $E$ and $L$, we obtain
\begin{eqnarray}\label{25}
E&=&\frac{a\mu L
\pm\sqrt{r^{2(D-3)}L^2-r^{D-3}\delta[r^{D-3}(r^2+a^2)+
a^2\mu]}\sqrt{\Delta}}{r^{D-3}(r^2+a^2)+a^2\mu},
\\\label{26}
L&=&\frac{-a\mu\pm\sqrt{r^{2(D-3)}E^2+(r^{D-3}-\mu)\delta
r^{D-3}}\sqrt{\Delta}}{r^{D-3}-\mu},
\end{eqnarray}
where we have used the following identity
\begin{eqnarray}\label{27}
r^2\Delta-a^2\mu^2&=&[r^{D-3}(r^2+a^2)+a^2\mu](r^{D-3}-\mu).
\end{eqnarray}
Equation (\ref{25}) can describe conditions for which $E$ can be
negative (seen by an observer at infinity).

First we assign $E=1$ to the particle at rest at infinity with unit
mass \cite{23}. We consider $+$ sign of Eq.(\ref{25}) which requires
$L<0$ for $E<0$ and
\begin{eqnarray}\nonumber
&&a^2L^2\mu^2
>\Delta[L^2r^{D-3}-(r^{2(D-3)}(r^2+a^2)+a^2\mu r^{D-3})]\delta.
\end{eqnarray}
Using Eq.(\ref{27}), this can be written as
\begin{equation}\nonumber
[r^{2(D-3)}(r^2+a^2)+a^2\mu r^{D-3}]
[L^2(1-\frac{\mu}{r^{D-3}})-\delta\Delta]<0.
\end{equation}
It follows from the above equation that $E<0$ if and only if $L<0$
and
\begin{equation}\nonumber
(r^{D-3}-\mu)<\frac{r^{D-3}\Delta}{L^2}\delta.
\end{equation}
We conclude that in higher dimensions only particles with
retrograde motion can have negative energy on the equatorial plane.
Also, it is necessary that particle remains inside the ergosphere.

\subsection{The Original Penrose Process}

The Penrose process describes that the particle with positive energy
enters into the ergosphere and breaks up into two parts such that
one will have negative while other part will have positive energy.
The particle with negative energy will be absorbed by the BH and the
particle with positive energy will escape to infinity. The particle
leaving the ergosphere will have more energy than the original
particle. The whole process results in decreasing the mass as well
as angular momentum of the BH. Hence, the rotational energy is
extracted from the BH in this process \cite{35}. Here, we suppose
that the photon absorbed by the BH (by crossing the event horizon)
possesses negative while the photon that escapes to infinity has
energy exceeding the original particle (which came from infinity).
Let $ E^{(1)}=1,~L^{(1)},~E^{(2)},~L^{(2)},~E^{(3)},~ L^{(3)}$ be
the energies and angular momenta of the original particle arrived
from infinity and the two photons (one that enters the event horizon
and other which escapes to infinity). The angular momentum of the
particle arrived from infinity by timelike geodesics can be obtained
from Eq.(\ref{26}) by setting $E=1$ and $\delta=-1$
\begin{equation}\label{28}
L^{(1)}=\frac{-a\mu+\sqrt{\mu r^{D-3}}\sqrt{\Delta}}
{r^{D-3}-\mu}=\sigma^{(1)}.
\end{equation}
The relationship between energies and angular momenta of the photon
that crosses the event horizon and the other which escapes to
infinity can be obtained by setting $\delta=0$ and considering both
negative and positive signs in Eq.(\ref{26}) as
\begin{eqnarray}\label{29}
L^{(2)}&=&\frac{[-a\mu-\sqrt{\Delta}r^{D-3}]E^{(2)}}
{r^{D-3}-\mu}=\sigma^{(2)}E^{(2)},
\\\label{30}
L^{(3)}&=&\frac{[-a\mu-\sqrt{\Delta}r^{D-3}]E^{(3)}}
{r^{D-3}-\mu}=\sigma^{(3)}E^{(3)}.
\end{eqnarray}

The conservation of energy and angular momentum yield
\begin{eqnarray}\nonumber
E^{(2)}+E^{(3)}=E^{(1)}=1,\quad
L^{(2)}+L^{(3)}=\sigma^{(2)}E^{(2)}+\sigma^{(2)}E^{(3)}=L^{(1)}=\sigma^{(1)},
\end{eqnarray}
which implies that
\begin{eqnarray}\nonumber
E^{(2)}=\frac{\sigma^{(1)}-\sigma^{(3)}}{\sigma^{(2)}-\sigma^{(3)}},
\quad
E^{(3)}=\frac{\sigma^{(2)}-\sigma^{(1)}}{\sigma^{(2)}-\sigma^{(3)}}.
\end{eqnarray}
Inserting $\sigma^{(1)},~\sigma^{(2)}$ and $\sigma^{(3)}$ from
Eqs.(\ref{28})-(\ref{30}), we obtain
\begin{eqnarray}\nonumber
E^{(2)}=-\frac{1}{2}\left[\sqrt{\frac{\mu}{r^{D-3}}}-1\right], ~
E^{(3)}=\frac{1}{2}\left[\sqrt{\frac{\mu}{r^{D-3}}}-1\right].
\end{eqnarray}
The photon that escapes to infinity has more energy than the
original particle $E^{(1)}=1$. Thus the gained energy ($\Delta E$)
can be written as
\begin{equation}\nonumber
\Delta
E=\frac{1}{2}\left[\sqrt{\frac{\mu}{r^{D-3}}}-1\right]=-E^{(2)}.
\end{equation}
According to Penrose process, the particle arriving from infinity
can attain maximum gain in energy at the event horizon. Thus
\begin{equation}\nonumber
\Delta
E\leq\frac{1}{2}\left[\sqrt{\frac{\mu}{r^{D-3}_{h+}}}-1\right].
\end{equation}
The maximum energy gain by the Kerr BH (in extreme limit) can be
achieved for $D=4$, i.e., $\Delta E=0.207$. The energy gain for MP
BH can be seen in Figure \textbf{2}. It is found that for $D=4$ we
have $\Delta E$ for the Kerr BH. For $D=5$, the energy gain is
higher than all dimensions while for $D=6$, it has almost similar
behavior as the Kerr BH. We have shown the energy gain for the
particles with positive spin as the behavior for the negative spin
remains the same due to symmetry. In all (eleven) dimensions, the
energy gain has increasing behavior but its value vary with respect
to dimensions.
\begin{figure}\centering
\epsfig{file=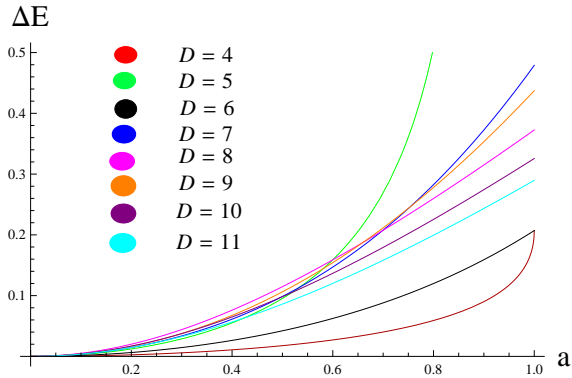,width=.55\linewidth} \caption{Energy gain versus
a.}
\end{figure}

\section{Center of Mass Energy in Higher Dimensions}

The rotating BHs have many interesting features, such as their
effects on the frequency shift on the photons that are traveling
near them as well as the extraction of rotational energy as the
photon with positive energy enters into the static limit of the BH.
One of the important features of rotating BHs is that they can also
act as particle accelerators. In this section, we study the center
of mass energy ($E_{cm}$) of the accelerating particles near MP BH
in the equatorial plane. The center of mass energy is defined as the
sum of rest masses and their kinetic energies of the two colliding
particles. It depends upon the nature of interacting particles as
well as astrophysical objects (BH or naked singularity) and
gravitational field surroundings such objects. It is interesting to
study the collision of particles as it is a naturally occurring
process in the universe. We consider two neutral colliding particles
having the rest masses $m_{1}$ and $m_{2}$. The conserved energy and
angular momentum of two particles are $E_{1}$, $E_{2}$, $L_{1}$ and
$L_{2}$. The angular momentum of the $i^{th}$ particle is defined as
\begin{equation}\label{a}
 k^{\nu}_{i}=m_{i}u^{\nu}_{i}, \quad i=1,2.
\end{equation}
The center of mass energy of the colliding particles is given as
\begin{equation}\label{aa}
E^2_{cm}=-k^{\nu}_{i}k_{i\nu}.
\end{equation}
Using Eq.(\ref{a}) in (\ref{aa}), we obtain
\begin{equation}\nonumber
\frac{E_{cm}}{\sqrt{2m_{1}m_{2}}}=\sqrt{\frac{(m_{1}-
m_{2})^2}{2m_{1}m_{2}}+(1-g_{\nu\eta}u^{\nu}_{1}u^{\eta}_{2})}.
\end{equation}
Substituting the values of $g_{\nu\eta}$, $u_{1}^{\nu}$ and
$u_{2}^{\eta}$, the center of mass energy in $D$-dimensions becomes
\begin{eqnarray}\nonumber
\frac{E_{cm}}{\sqrt{2m_{1}m_{2}}}=[\frac{(m_{1}-
m_{2})^2}{2m_{1}m_{2}}+\frac{1}{r^{D-3}\Delta}[r^{D-3}\Delta+
E_{1}E_{2}(r^{D-1}+a^2r^{D-3}+a^2\mu)
\\\nonumber
-(E_{2}L_{1}+E_{1}L_{2})(a\mu)-L_{1}L_{2}(r^{D-3}-\mu)-\sqrt{R_{1}R_{2}}]]^{\frac{1}{2}},
\end{eqnarray}
where
\begin{equation}\nonumber
R_{i}=r^{D-1}E^2_{i}+\mu(aE_{i}-L_{i})^2-r^{D-3}L^2_{i}-r^{D-1}+\mu
r^2, \quad i=1,2.
\end{equation}
For the sake of simplicity, we take $m_{1}=m_{2}=m_{0}$ and
$E_{1}=E_{2}=E=1$ such that the above equation takes the form
\begin{eqnarray}\nonumber
\frac{E_{cm}}{\sqrt{m^2_{0}}}&=&[\frac{1}{r^{D-3}\Delta}[r^{D-3}\Delta+
(r^{D-1}+a^2r^{D-3}+a^2\mu)
\\\nonumber
&-&(L_{1}+L_{2})(a\mu)-L_{1}L_{2}(r^{D-3}-\mu)-\sqrt{R_{1}R_{2}}]]^{\frac{1}{2}},
\end{eqnarray}
where
\begin{equation}\nonumber
R_{i}=\mu(a-L_{i})^2-r^{D-3}L^2_{i}+\mu r^2, \quad i=1,2.
\end{equation}
For $D=4$, the above equation reduces to \cite{27}. Figure
\textbf{2} shows the behavior of the center of mass energy for
$M=1,L_{1}=2,L_{2}=2.5$ upto eleven dimensions. We found that the
center of mass energy has decreasing behavior in all dimensions.
\begin{figure}\centering
\epsfig{file=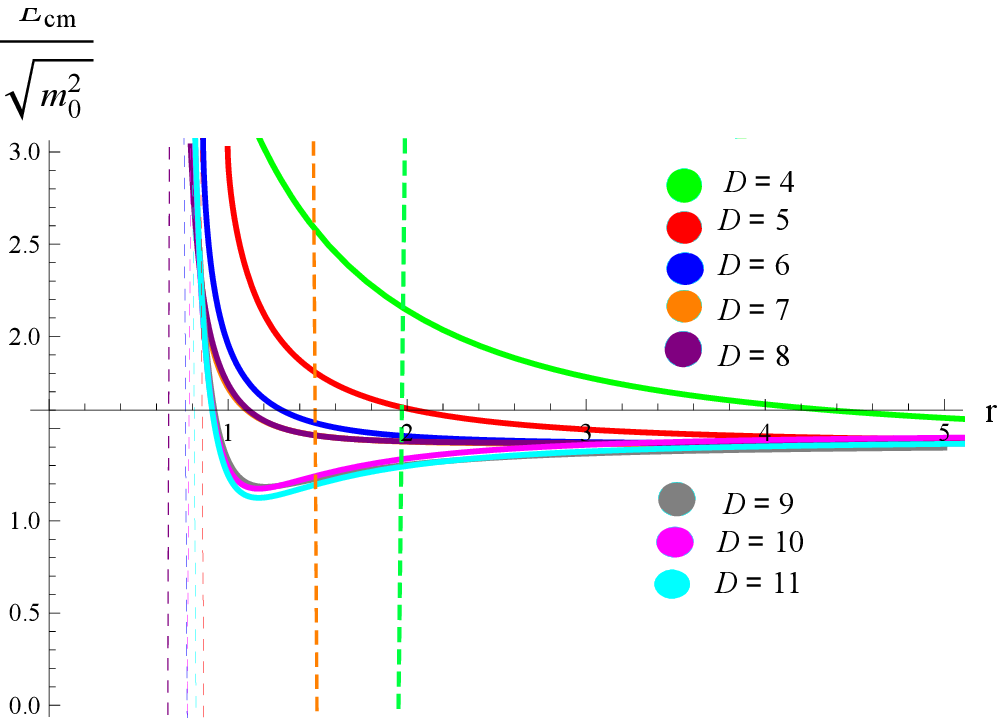,width=.52\linewidth}\epsfig{file=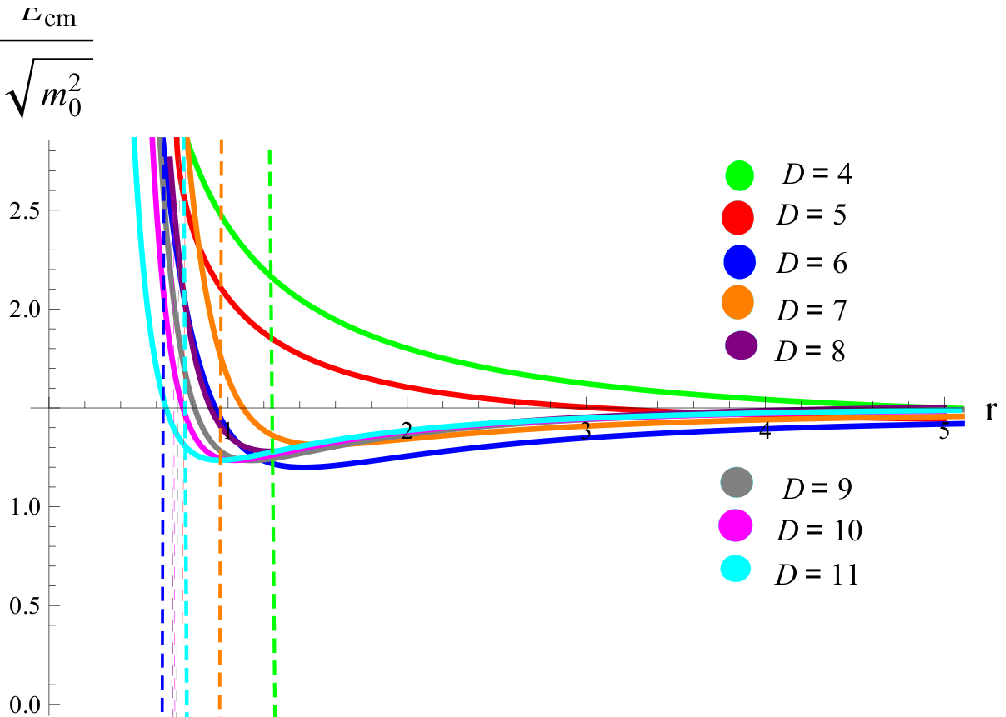,width=.52\linewidth}
\caption{Center of mass energy with respect to $r$ for $a=0.25$
(left) and $a=0.6$ (right). Here, the curves show
$E_{cm}/\sqrt{m^2_{0}}$ while vertical dashed lines are event
horizons.}
\end{figure}

\section{Final Remarks}

When a light wave (from an approaching galaxy which is moving
towards an observer) gets scrunched to the shorter wavelength, this
is known as galactic blue shift. On the other hand, if a light wave
from a galaxy moving away from the observer gets stretched to the
longer wavelength then this is called galactic red shift. Slipher
discovered that the Andromeda galaxy possesses a large blue shift
which indicates that this galaxy moves towards the Earth. He further
investigated other spiral galaxies and found that most of them have
large red shift indicating that they are moving away from us. Hubble
observations indicate that relative to the Earth and all the
observed galactic objects, galaxies are receding in every direction.
The velocities calculated from their observed red shifts are
directly proportional to their distance from each other as well as
from the Earth. Hubble was the pioneer in explaining the expanding
universe with red shift phenomena \cite{36}.

It is well-known that there is a supermassive BH (SgrA*) in the
center of Milky Way as well as in many other spiral galaxies. In
this paper, we assume higher dimensional MP BH as supermassive BH at
the galactic center. Motivated by \cite{21,22}, we generalize the
mass and angular momentum parameter in arbitrary extra dimensions in
terms of red-blue shifts of the photons emitted from circular
timelike geodesic and traveling along the null geodesic. For this
purpose, we have first calculated red-blue shifts of the photons in
higher dimensions for an observer located far away. We have taken
circular as well as equatorial orbits to find these shifts. We have
expressed the corresponding mass, rotation parameter and radius of
the detector in terms of red-blue shifts. In this way, we have
generalized the results for the Kerr BH. We have only discussed the
analytical model, however, these results may be useful if they can
be calculated using the observational data. The generalized results
can provide information about the behavior of BH parameters in
dimensions $D\geq4$.

It is believed that the supermassive BHs (powering the active
galaxies and quasars) are the rapidly rotating BHs. Such BHs produce
powerful jets of gas (whose direction is sometimes stable over
million of years) whose source of energy may be the rotation of BH.
It may be possible that this rotational energy is extracted due to
Penrose process \cite{35}. Following \cite{23}, we have studied the
Penrose process for the MP BH. We have found that particle will have
negative energy for the retrograde motion ($L<0$) in higher
dimensions. We have also seen that the energy gain of the particle
is dimension dependent. The energy gain for $D=4$ and $D=6$ have the
same value while for $D=5$, it has the highest value.

We have also examined the influence of higher dimensions on the
center of mass energy of two colliding particles. We have plotted
$E_{cm}$ by considering two different values of the rotating
parameters. The center of mass energy decreases with increasing
radius. For $a=0.25$, the center of mass energy for $D=6~\text{and}~
8$, it lies inside the event horizon, while for other dimensions, it
crosses the event horizon. For $a=0.6$, the center of mass energy
crosses the event horizon in all dimensions. In both cases, the
center of mass energy for $D=4$ is greater than that of $D\geq5$.
Finally, we conclude that the motion of particles in higher
dimensions experiences a very different behavior than the four
dimensions.

\vspace{0.5cm}

{\Large \bf Dedication}

\vspace{0.5cm}

\textbf{We would like to dedicate this paper to Prof. Asghar Qadir
on his 70th birthday (July 23). May he has a healthy long life!}

\end{document}